\documentclass[twocolumn,prb,showpacs,superscriptaddress]{revtex4-1}

\usepackage{psfrag}
\usepackage{graphicx}
\usepackage{amsmath}
\usepackage{amssymb}
\usepackage{color}

\begin{document}
\title{Dzyaloshinskii-Moriya interaction induced extrinsic linewidth broadening of ferromagnetic resonance}
\author{Jung Hyun Oh}
\affiliation{Department of Materials Science and Engineering, Korea University, Seoul 02841, Korea }
\author{Ik-Sun Hong}
\affiliation{KU-KIST Graduate School of Conversing Science and Technology, Korea University, Seoul 02841, Korea}
\author{Kyung-Jin Lee}\email{kj\_lee@korea.ac.kr}
\affiliation{Department of Materials Science and Engineering, Korea University, Seoul 02841, Korea }
\affiliation{KU-KIST Graduate School of Conversing Science and Technology, Korea University, Seoul 02841, Korea}

\date{\today}

\begin{abstract}
{
For a thin ferromagnetic film with the Dzyaloshinskii-Moriya interaction (DMI),
we derive an expression of the extrinsic ferromagnetic resonance (FMR) linewidth in a quantum mechanical way,
taking into account scatterings from structural inhomogeneity.
In the presence of the DMI, the magnon dispersion exhibits rich resonant states, especially 
in small external magnetic fields and strong DMI strength. It is found that the FMR linewidth shows several characteristic features such as a finite linewidth at zero frequency and peaks in the low frequency range.
}
\end{abstract}

\maketitle 

\section{Introduction}

Magnetic damping~\cite{Kambersky1976,Gilmore,Kambersky2007,Garate,Ebert,Yuan,Schoen}, which is parameterized by the Gilbert damping constant $\alpha$, describes the energy dissipation rate of magnetization dynamics and determines the performance of magnetic devices including magnetic random access memories and magnetic sensors~\cite{Sun,SWL,Smith,Safonov}. The ferromagnetic resonance (FMR) provides a way to measure the damping $\alpha$ from the FMR linewidth~\cite{Mizukami,Kobayashi,Langner,Liu,Weindler}. A widely adopted method for the estimation of $\alpha$ is to measure the frequency-dependent change in the linewidth, assuming that the only intrinsic damping contribution is proportional to the frequency.  However, this method is not always straightforward as the measured linewidth includes not only intrinsic damping contribution but also extrinsic contributions originating from inhomogeneity of the sample. Especially, the contribution of two-magnon scattering~\cite{Sparks,Celinski,Hurben,McMichael,Arias,McMichael1,McMichael2,Zakeri,Landeros} to the linewidth also depends on the frequency, demanding a detailed understanding of the frequency-dependent linewidth of two-magnon scattering origin~\cite{Arias,Landeros}.

Recently, there has been much interest in the spin-orbit coupling effect for a thin ferromagnetic layer in contact with a heavy metal layer because it permits the control and manipulation of electronic spin degree of freedom~\cite{Miron,Liu2,Emori,Ryu,Haney1,Haney2,Fan,Qiu,Oh,Amin1,Amin2,Humphries,Baek,Kim}.
In the presence of spin-orbit interaction combined with inversion symmetry breaking, the antisymmetric exchange interaction, known as Dzyaloshinskii-Moriya interaction (DMI), emerges~\cite{Dzyaloshinskii,Moriya}
in addition to the symmetric exchange interaction responsible for ferromagnetism.
The DMI affects equilibrium spin textures by stabilizing chiral orders such as spin spirals, chiral domain walls, and magnetic skyrmions~\cite{Rossler2006,Uchida,Huang,Fert1980,Bode2007,Heide2008,Udvardi2009,Zakeri2010,Costa,Kim2,Thiaville,Chen2013,Muhlbauer2009,
Yu2010Nat,Jiang2015,Woo,Moreau,Boulle,Pollard} because the DMI energetically favors an orthogonal configuration of two neighboring spins with a fixed rotating sense. The DMI also affects spin wave properties by differentiating the right- and left-moving spin waves because of their different chirality~\cite{Cort,Moon,Di,Nembach,Cho,Lee2016,Seki,JV,Lan,Yu2016,Lan2,LSJ}. In this respect, it is important to investigate the contribution of DMI inhomogeneity to the extrinsic broadening of FMR linewidth through two-magnon scattering process.

In this work, we theoretically investigate the FMR linewidth of a thin ferromagnetic layer with DMI in a quantum mechanical way.
A two-magnon model is used to describe an extrinsic source for the FMR linewidth,
taking into account scatterings caused by structural inhomogeneities.
The inhomogeneity may arise from a wide variety of microstructural origins.
Here, we consider atomic-sized structural imperfections and associated fluctuation of DMI strength in space.

\section{Calculational model}

\begin{figure}[b]
\centering
\includegraphics[width=45.0ex, angle=0.0]{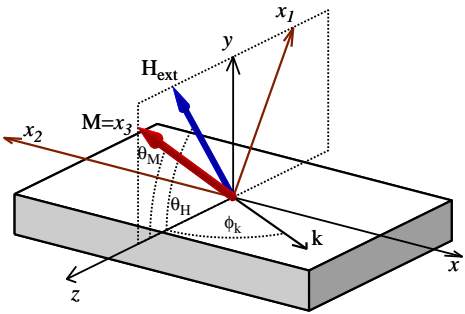}
\caption{(color online) Schematic representation of a ferromagnetic thin film. We choose
a coordinate system for applied magnetic field and 
equilibrium  magnetization to be in the $yz$-plane. 
A local coordinate system ${\bf {\hat x}}_{1,2,3}$ is shown, where
the ${\bf {\hat x}}_{3}$ axis is coincide to the equilibrium magnetization direction.
\label{devstr}
}
\end{figure}

We consider a thin ferromagnetic film with the thickness $d$ as illustrated in Fig.~\ref{devstr}. 
We chose the film surface parallel to the $zx$ plane and the $y$-axis normal to the surface.
An external magnetic field $H_{\rm ext}$ is applied
in the $yz$ plane with an angle $\theta_H$ from the $z$-axis.
Then, the equilibrium magnetization is also in that plane with a tilt angle $\theta_M$, determined by the competition of magnetic energies.

We assume that the magnetization spatially varies in the $zx$-plane while it is uniform in the $y$-direction due to the small thickness.
Thus, we describe magnon modes with a wave vector,
${\bf k}= k(\cos\phi_{\bf k} {\bf {\hat e}}_z +\sin\phi_{\bf k}  {\bf {\hat e}}_x)$ with a wave vector modulus, $k$.
In this case, a demagnetization factor of the system is given by $N_{\rm d} = (1-e^{-kd})/kd$~[\onlinecite{DE}].
We also make use of the ${\bf {\hat x}}_{1,2,3}$ coordinate system.
As shown in Fig.~\ref{devstr}, the ${\bf {\hat x}}_3$ axis is coincided with
the equilibrium magnetization direction while the ${\bf {\hat x}}_2$ axis is directed to
the negative $x$-axis.
Thus, $H_1({\bf k}) = \cos\theta_M H_{\rm ext}^y({\bf k})$
and $H_2({\bf k}) =-H_{\rm ext}^x({\bf k})$.

To be specific, the Hamiltonian ${\cal H}_0$ consists of the demagnetization, exchange, anisotropy, and Zeeman energies;
${\cal H}_0 = {\cal H}_{\rm demag} + {\cal H}_{\rm ex} + {\cal H}_{\rm A}+{\cal H}_{\rm Z}$.
In a second-quantized form, ${\cal H}_0$ is summarized as~\cite{Kalinikos, McMichael, McMichael1}
\begin{eqnarray}
 {\cal H}_0 &=& \sum_{\bf k} \left[ \sum_{\kappa\kappa'=1,2}  m_\kappa(-{\bf k})
 \epsilon_{\kappa \kappa'}({\bf k}) m_{\kappa'}({\bf k}) \right.
\nonumber\\
&&\left.-\sqrt{2\hbar\gamma M_S}\sum_{\kappa=1,2}  m_\kappa({\bf k}) 
\mu_0 H_\kappa(-{\bf k}) \right],
\label{H0}
\end{eqnarray}
where 
\begin{eqnarray}
\epsilon_{11} &=& \epsilon^0_{11}\!+\!\epsilon_{\rm M}( \sin^2\theta_M\cos^2\phi_{\bf k}\!-\! \cos 2\theta_M )
\!+\!\epsilon_{\rm ex} k^2d^2,
\nonumber\\
\epsilon_{22} &=& \epsilon^0_{22}\!+\!\epsilon_{\rm M}(\sin^2\phi_{\bf k}\!+\!\sin^2\theta_M )
\!+\!\epsilon_{\rm ex} k^2d^2,
\nonumber\\
\epsilon_{12} &=& \epsilon_{\rm M} \sin\phi_{\bf k}\cos\phi_{\bf k} \sin\theta_M
\!+\!i \epsilon_{\rm DM} kd\sin\phi_{\bf k}\cos\theta_M,
\nonumber\\
\epsilon_{21} &=& \epsilon_{12}^*.
\label{epsilons}
\end{eqnarray}
Here, we define $\epsilon_{\rm M} = \hbar\gamma \mu_0 M_{\rm S}(1\!-\!N_{\rm d})$ 
with the saturation magnetization $M_{\rm S}$ and the gyromagnetic ratio $\gamma$,
$\epsilon_{\rm ex}={2\hbar\gamma A}/{M_{\rm S}d^2}$
with the exchange stiffness constant $A$, and
$\epsilon_{\rm DM}={4\hbar\gamma D}/{M_{\rm S} d}$
with the DMI strength $D$.
$\epsilon^0_{11}$ and $\epsilon^0_{22}$ are ${\bf k}$-independent terms
and originate from both Zeeman and surface anisotropy energies
with an anisotropy field $H_S$;
\begin{eqnarray}
\epsilon^0_{11} &=& \hbar\gamma \mu_0\left[ H_3 +(M_{\rm S}\!+\!H_S) \cos 2\theta_M \right],
\\
\epsilon^0_{22} &=& \hbar\gamma\mu_0\left[ H_3 -(M_{\rm S}\!+\!H_S)\sin^2\theta_M \right].
\label{energy0}
\end{eqnarray}
where
$H_3=H_{\rm ext}\cos(\theta_H\!-\!\theta_M)$ is the ${\bf {\hat x}}_3$-component of magnetic field $H_{\rm ext}$.  By requiring the $M_x=0$ in equilibrium, the magnetization
angle $\theta_M$ is determined by $2H_{\rm ext}\sin(\theta_H\!-\!\theta_M)=(H_S\!+\!M_{\rm S})\sin 2\theta_M$
for a given magnetic field.

We express transverse components of the magnetization,
$M_{1,2}=\sqrt{2\hbar\gamma M_S} m_{1,2}$ with bosonic annihilation
and creation operators approximated from the Holstein-Primakoff setup~\cite{Holstein};
\begin{eqnarray}
m_1({\bf k}) &=& \frac{1}{2}\left(c_{\bf k}+c_{-{\bf k}}^\dagger \right),
\\
m_2({\bf k}) &=& \frac{1}{2i}\left(c_{\bf k}-c_{-{\bf k}}^\dagger \right), 
\end{eqnarray}
satisfying usual bosonic commutation relations of
$[c_{\bf k},c_{\bf k}]=0$,
$[c_{\bf k}^\dagger,c_{\bf k}^\dagger]=0$,
and $[c_{\bf k},c_{\bf k}^\dagger]=\delta_{{\bf k},{\bf k}'}$.

Eigenenergy of magnon modes can be obtained by the Bogoliubov transformation of
the Hamiltonian ${\cal H}_0$. In terms of $\epsilon_{\kappa\kappa'}$,
the eigenenergy at a ${\bf k}$ point is given by,
\begin{equation}
\epsilon({\bf k}) = {\rm Im}[\epsilon_{12}({\bf k})]+\sqrt{\epsilon_{11}({\bf k})\epsilon_{22}({\bf k})
-{\rm Re}[\epsilon_{12}({\bf k})]^2 }.
\label{eigenenergy}
\end{equation}

\begin{figure}[ht]
\centering
\includegraphics[width=40.0ex, angle=0.0]{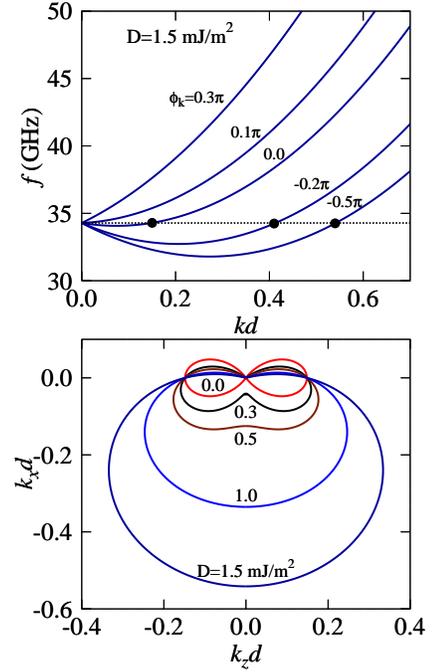}
\caption{(color online) Magnon dispersion of a ferromagnetic thin film in the presence of DMI ($D=1.5 $mJ/m$^2$) is
plotted for various ${\bf k}$ directions (upper panel).
Equi-energy lines at a fixed energy equal to $\epsilon({\bf k}\!=\!0)$ are plotted in the $(k_z,k_x)$ plane
 for several DMI strengths (lower panel).
Used parameters are $d=5$nm, $A=1.3\times 10^{-11}$J/m, $\mu_0M_{\rm S}=1.0$ T, $\mu_0 H_S=-0.5 $T,
and $\mu_0 H_{\rm ext}=1.0$ T with $\theta_H=0$.
\label{fig_disp} }
\end{figure}

For two-magnon scattering of FMR experiment, we focus on the energy $\epsilon({\bf k}=0)=\sqrt{\epsilon_{11}^0\epsilon_{22}^0}$
and its resonant states, namely a set of states at ${\bf k}\neq 0$ with the same energy.
We plot the dispersion of the system at various DMI strengths in the upper panel of Fig.~\ref{fig_disp}.
In the figure, we display those resonant states with solid circles, satisfying the relation of $\epsilon({\bf k}) =\sqrt{\epsilon_{11}^0\epsilon_{22}^0}$.
We find that the calculated dispersion is different from the DMI-free case
mainly in two points; the resonant state with the the largest $|{\bf k}|$
occurs around  the angle $\phi_{\bf k}=-\pi/2$ (in the case of $D<0$, $\phi_{\bf k}=\pi/2)$ and moreover,
its value of $kd$ is not small,
but can be comparable to one. 
In the absence of DMI, resonant states are determined from
the competition between exchange and demagnetization energies. In this case,
the resonant state with the largest $k$ occurs at $\phi_{\bf k}=0$. In the presence of DMI, on the other hand, an additional decreasing energy is incorporated
in the region of $k\sin\phi_{\bf k}\!<\!0$ through $\epsilon_{12}({\bf k})$ of Eq.~(\ref{epsilons}).
Consequently, as $k$ increases, this energy competes with an increasing part (i.e., the exchange energy) and eventually, the largest-$k$ resonant state
is formed around $\phi_{\bf k}=-\pi/2$. 

In order to show the tendency clearly, we plot the evolution of resonant states
in the lower panel of Fig.~\ref{fig_disp} for several DMI strengths.
In a DMI-free case ($D=0$), one finds two lobes having the largest distances at $\phi_{\bf k}=0$ and $\pi$, respectively.
As we increase DMI strength, 
the two lobes become one at a certain DMI energy (say $\epsilon_{\rm DM}^0$) by creating a resonant state at $\phi_{\bf k}=-\pi/2$.
Eventually, for a sufficiently strong DMI, the resonant state with the largest $k$ appears
at $\phi_{\bf k}=-\pi/2$.

According to the two-magnon theory, inhomogeneities allow a magnon mode at ${\bf k}=0$
to scatter into modes with finite wave vectors and couple these modes with different wave vectors. 
To describe this coupling, we add a perturbation term into the Hamiltonian as, ${\cal H}={\cal H}_0+{\cal H}_{\rm pertub}$;
\begin{equation}
{\cal H}_{\rm pertub} = \sum_{{\bf k},{\bf k}'} \sum_{\kappa,\kappa'=1,2} m_\kappa(-{\bf k})
U_{\kappa\kappa'}^{{\bf k},{\bf k}'} m_{\kappa'}({\bf k}').
\label{Hpertub}
\end{equation}
In general, the potential $U_{\kappa\kappa'}^{{\bf k},{\bf k}'}$ is not diagonal
in the ${\bf k}$-space and thus causes a coupling between different ${\bf k}$-states.
The perturbation potential from inhomogeneities is zero in average, but it provides
appropriate momentum and energy for magnon energy levels to be shifted and broaden.

In this work, we focus on DMI-induced perturbations.
Inferring from the imaginary part of $\epsilon_{12}({\bf k})$ in Eq.~(\ref{epsilons}) relevant to
the DMI, we consider two feasible mechanisms as perturbation sources.
One is a random spatial variation of the parameter $D$ 
while the other is an abrupt change of the magnetization vectors, for example, at edges of terrace on sample surfaces,
near magnetic and non-magnetic impurities, etc.
If structural defects whose characteristic size is larger or equal to the film thickness $d$,
the magnetization will be slowly varying and the former is dominant for the fluctuation, whereas, in a scale much less than $d$, the latter is dominant. 
In both cases, the fluctuation potential is modeled as,
\begin{eqnarray}
U_{11}^{{\bf k},{\bf k}'} &=& ~U_{22}^{{\bf k},{\bf k}'} = 0,
\nonumber\\
U_{12}^{{\bf k},{\bf k}'} &=&-U_{21}^{{\bf k}',{\bf k}} = 
\sum_j e^{-i({\bf k}\!-\!{\bf k}')\cdot {\bf R_j}} U^a({\bf k},{\bf k}'),
\end{eqnarray}
where ${\bf R}_j$ is a random position vector of impurity (localized inhomogeneity).
By defining $\lambda_c$ as a characteristic length scale of the impurity,
we model its localized potential with Fourier components of
 (see Appendix A for details),
\begin{eqnarray}
U^a({\bf k},{\bf k}') \!=\! \epsilon_{\rm DM}\!
f_R({\bf k}\!-\!{\bf k}')
\!\cos\theta_M {\cal W}({\bf k},{\bf k}'),
\label{Ua}
\end{eqnarray}
where 
${\cal W}({\bf k},{\bf k}') = i (k'_x+k_x) d/2$ for $\lambda_c \gtrsim d$, whereas ${\cal W} = d/\lambda_c$ for $\lambda_c \ll d$. Here, 
$f_R({\bf k})$ is a form factor of the localized potential; for example,
$f_R({\bf k})= e^{-k^2\lambda_{\rm imp}^2} \pi\lambda_{\rm imp}^2/L^2$
if we assume Gaussian shapes with an impurity size $\lambda_{\rm imp}$
distributed in the $(L\times L\times d)$-sized ferromagnetic film.

\section{ Susceptibility }

We shall calculate FMR linewidth by examining the susceptibility of the system.
According to equation of motion for $m_{1,2}({\bf k},t)$, the 
susceptibility for an external field with wave-number ${\bf k}'$ and time-varying $e^{i\omega t}$ is given by,
\begin{eqnarray}
\left( \begin{array}{cc}
\chi_{12} & \chi_{11} \\
\chi_{22} & \chi_{21} \\
\end{array} \right) ({\bf k},{\bf k}';\omega) = [{\cal G}^{-1}\delta_{{\bf k},{\bf k}'}-{\cal V}]^{-1},
\end{eqnarray}
where
\begin{eqnarray}
{\cal G}({\bf k};\omega) &=& 
\hbar\gamma M_{\rm S}
\left( \begin{array}{cc}
-i\hbar\omega+\epsilon_{21}({\bf k}) & \epsilon_{22}({\bf k}) \\
\epsilon_{11}({\bf k}) & i\hbar\omega+\epsilon_{12}({\bf k}) \\
\end{array} \right)^{-1},
\nonumber\\
{\cal V}({\bf k},{\bf k}') &=& 
-\frac{1}{\hbar\gamma  M_{\rm S}}
\left( \begin{array}{cc}
V_{21}({\bf k},{\bf k}') & V_{22}({\bf k},{\bf k}') \\
V_{11}({\bf k},{\bf k}') & V_{12}({\bf k},{\bf k}') \\
\end{array} \right)
\end{eqnarray}
with $V_{\kappa\kappa'}({\bf k},{\bf k}') = (U_{\kappa\kappa'}^{{\bf k},{\bf k}'}
+U_{\kappa'\kappa}^{-{\bf k}',-{\bf k}} )/2$.

\subsection{ Average over impurities }

In order to  average the susceptibility over an ensemble of impurity, 
we expand the susceptibility with perturbation potential ${\cal V}$;
\begin{eqnarray}
\left( \begin{array}{cc}
\chi_{12} & \chi_{11} \\
\chi_{22} & \chi_{21} \\
\end{array} \right)  &=&
 {\cal G}+{\cal G}\left\langle {\cal V}\right\rangle_{\rm imp}{\cal G}
+{\cal G}\left\langle{\cal V}{\cal G}{\cal V}\right\rangle_{\rm imp}{\cal G}+\cdots
\nonumber\\
&=& [{\cal G}^{-1}\delta_{{\bf k},{\bf k}'}-{\cal C}]^{-1}.
\label{chi_avg}
\end{eqnarray}
where $\langle\dots\rangle_{\rm imp}$ denotes an average over impurity and
usually, 
$\left\langle {\cal V}\right\rangle_{\rm imp}=0$ due to the randomness of 
${\cal V}$. We approximate a self-energy as 
${\cal C}=\left\langle{\cal V}{\cal G}{\cal V}\right\rangle_{\rm imp}$ by neglecting crossing terms.
Because the fluctuating potential is contributed from randomly located sites at $\{{\bf R}_j\}$
and thus $V_{\kappa\kappa'}({\bf k},{\bf k}') = \sum_j V^a_{\kappa\kappa'}({\bf k},{\bf k}')e^{-i({\bf k}-{\bf k}')\cdot{\bf R}_j}$,
we arrive at the self-energy of
\begin{flalign}
&{\cal C}({\bf k},{\bf k}';\omega) = 
\delta_{{\bf k},{\bf k}'}
\frac{1}{\hbar\gamma  M_{\rm S}}
 \left( \begin{array}{cc}
\Sigma_{21} & \Sigma_{22} \\
\Sigma_{11} & \Sigma_{12} \\
\end{array} \right) ({\bf k};\omega)
\end{flalign}
where
\begin{flalign}
\left( \begin{array}{cc}
\Sigma_{21} & \Sigma_{22} \\
\Sigma_{11} & \Sigma_{12} \\
\end{array} \right)
&({\bf k};\omega) =
\frac{ N_{\rm imp} }{\hbar\gamma M_s} 
\sum_{{\bf k}_1} 
\left( \begin{array}{cc}
V^a_{21} & V^a_{22} \\
V^a_{11} & V^a_{12} \\
\end{array} \right)
({\bf k},{\bf k}_1)
\nonumber\\
&{\cal G}({\bf k}_1;\omega)
\left( \begin{array}{cc}
V^a_{21} & V^a_{22} \\
V^a_{11} & V^a_{12} \\
\end{array} \right)
({\bf k}_1,{\bf k})
\label{SigmaImp}
\end{flalign}
with the number of impurities $N_{\rm imp}$.
We note that the self-energy of ${\cal C}({\bf k},{\bf k}';\omega)$ is now diagonal over the ${\bf k}$-space,
meaning that a translation symmetry is recovered through the impurity average.
Finally, the susceptibility is given as,
\begin{eqnarray}
\left( \begin{array}{cc}
\chi_{12} & \chi_{11} \\
\chi_{22} & \chi_{21} \\
\end{array} \right)&&({\bf k},{\bf k}';\omega)=
\hbar\gamma  M_{\rm S} 
\nonumber\\
&&\left( \begin{array}{cc}
-i\hbar\omega\!+\!{\tilde\epsilon}_{21}&{\tilde\epsilon}_{22} \\
{\tilde \epsilon}_{11} & 
\!i\hbar\omega\!+\!{\tilde \epsilon}_{12} \\
\end{array} \right)^{-1} \delta_{{\bf k},{\bf k}'}
\label{chifull}
\end{eqnarray}
with renormalized energy components of ${\tilde \epsilon}_{\kappa\kappa'}({\bf k},\omega)
=\epsilon_{\kappa\kappa'}({\bf k})-\Sigma_{\kappa\kappa'}({\bf k};\omega)$.

\subsection{ Linewidths }

Now we focus on the longitudinal susceptibility along the $x$-direction, $\chi_{xx}=\chi_{22}$, from Eq.~(\ref{chifull});
\begin{eqnarray}
\chi_{xx}({\bf k},{\bf k};\omega) &=&
\frac{ \hbar\gamma  M_{\rm S} {\tilde \epsilon}_{11} }{ 
[i\hbar\omega\!-\!{\tilde\epsilon}_{21}] [i\hbar\omega\!+\!{\tilde\epsilon}_{12}]
 +{\tilde\epsilon}_{11} {\tilde\epsilon}_{22} }.
\label{chixx}
\end{eqnarray}
In the absence of DMI, this result is consistent with that in Ref. [\onlinecite{Landeros}].
At this stage, we insert the intrinsic damping contribution by adding $i\hbar\omega \alpha$ to $\epsilon_{11}({\bf k})$ and $\epsilon_{22}({\bf k})$.
In FMR experiments, since the frequency $\omega$ is fixed to $\omega_{\rm FMR}$
and the dc field $H_{\rm ext}$ is swept,
a linewidth of the FMR resonance is equal to Lorentzian broadening
of Eq.~(\ref{chixx}) at ${\bf k}=0$.
A straightforward calculation leads to the linewidth of,
\begin{flalign}
\hbar \gamma &\mu_0\cos(\theta_M\!-\!\theta_H) \Delta H_{\rm ext}
=\alpha \hbar\omega_{\rm FMR} 
\nonumber\\
&+ \frac{ {\rm Im}[ \epsilon^0_{11}\Sigma_{22}({\bf k},\omega)\!+\!
\epsilon^0_{22}\Sigma_{11}({\bf k},\omega) ]^{{\bf k}=0}_{\omega=\omega_{\rm FMR}}
}{(\epsilon^0_{11}\!+\!\epsilon^0_{22})}.
\end{flalign}
Using the perturbation potential of Eq.~(\ref{Ua}),
this result can be summarized as
\begin{equation}
\hbar \gamma \mu_0 \Delta H_{\rm ext} = \frac{\alpha}{\cos(\theta_M\!-\!\theta_H)} \hbar\omega_{\rm FMR}
+\hbar\gamma \mu_0 \Delta H^{(2)},
\end{equation}
where $\Delta H^{(2)}$ stands for the extrinsic contribution from the two-magnon scattering.
Detailed forms of those quantities depend on the model of perturbation potential.
For instance, in the case of Eq.~(\ref{Ua}), the extrinsic contribution is given by,
\begin{equation}
\hbar\gamma \mu_0 \Delta H^{(2)}
= \left(
\frac{N_{\rm imp}}{L^2} \frac{(\pi\lambda^2_{\rm imp})^2}{\lambda_c^2 }
\frac{\epsilon_{\rm DM}^2}{\epsilon_{\rm ex}} \cos^2\theta_M \right) \Gamma(\omega).
\label{H2}
\end{equation}
Here, the amplitude (the part enclosed by the parenthesis) depends on the model of scattering potential
and a dimensionless function $\Gamma(\omega)$ is determined purely by the magnon dispersion and
a cut-off function through,
\begin{flalign}
\Gamma&(\omega)=
\frac{1}{\cos(\theta_M\!-\!\theta_H)}
\frac{\epsilon_{\rm ex} }{ (\epsilon_{22}^0\!+\!\epsilon^0_{11}) }
\frac{d^2}{L^2} {\rm Im} \sum_{{\bf k}} f_c(k)
\nonumber\\
&
\frac{\epsilon_{11}^0\epsilon_{11}({\bf k})+\epsilon_{22}^0\epsilon_{22}({\bf k}) }
{ [ i\hbar\omega\!-\!\epsilon_{21}({\bf k}) ]
[i\hbar\omega\!+\!\epsilon_{12}({\bf k})] +\epsilon_{11}({\bf k}) \epsilon_{22}({\bf k}) },
\label{Gamma}
\end{flalign}
where 
$f_c(k)=e^{-k^2\lambda_{\rm imp}^2} (k\lambda_c)^2$ for a large-sized impurity model ($\lambda_c\gtrsim d$)
and $f_c(k)=e^{-k^2\lambda_{\rm imp}^2}$ for a small-sized case ($\lambda_c\ll d$).
Other scattering models,
such as anisotropy-induced scattering potential by surface roughness, can be reproduced
by simply replacing $f_c(k)=1$, but with the modified amplitude
that includes detailed form-factors of impurity~\cite{Arias}.
In calculating the function $\Gamma(\omega)$,
the summation over ${\bf k}$ is mainly contributed by those ${\bf k}$-points where the denominator,
$ [ i\hbar\omega\!-\!\epsilon_{21}({\bf k}) ]
[i\hbar\omega\!+\!\epsilon_{12}({\bf k})] +\epsilon_{11}({\bf k}) \epsilon_{22}({\bf k})$,
approaches zero.
These ${\bf k}$-points correspond to the resonant states discussed in the previous section.
Hence, the total number of resonant state is important in determining 
$\Gamma(\omega)$ and eventually the FMR linewidth.

\section{ results and discussion }

Now we proceed to study behavior of the linewidth by varying the DMI strength $D$.
In order to present calculated results in a simple way, the perturbation potential model
by a small sized impurities of $\lambda_c \ll d$ is mainly discussed
for a fixed film thickness of $d=5$ nm. We note that $\Gamma(\omega)$ is much smaller for a large size impurity of $\lambda_c \gg d$. 
As an example, 
we choose parameters of impurity potential such as
impurity density of $N_{\rm imp}/L^2 = 0.03/(\pi\lambda_{\rm imp}^2)$,
$\lambda_c=1$ nm, and $\lambda_{\rm imp}=0.5$ nm.
For the exchange stiffness constant $A=1.3\times 10^{-11}$ J/m and DMI strength $D=1.5$ mJ/m$^2$,
the chosen parameters give $1.186 \cos^2\theta_M$ $\mu$eV for the amplitude of Eq.~(\ref{H2}),
which corresponds to $102.5 \cos^2\theta_M$ G with $\hbar\gamma=0.1158$ meV/T.

\begin{figure}[b]
\centering
\includegraphics[width=45.0ex, angle=0.0]{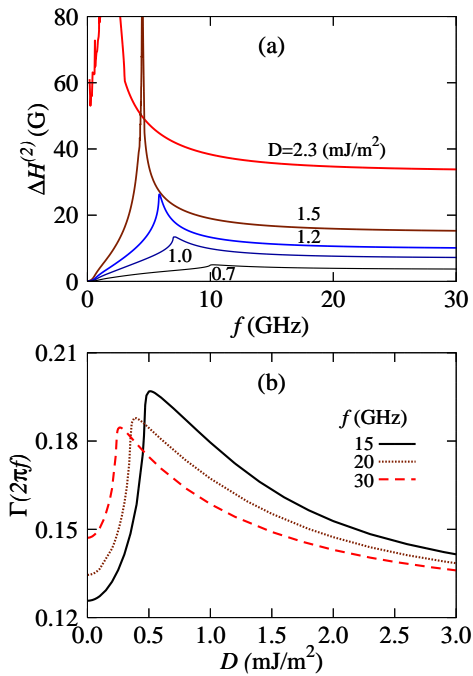}
\caption{(color online) In (a), the calculated linewidths for various DMI strengths $D$ are plotted as a function of FMR frequency. 
In (b), we show the variation of $\Gamma(\omega)$ with increasing DMI strength by fixing FMR frequency.
The exchange stiffness constant, $A=1.3\times 10^{-11}$ J/m is used and the external magnetic field
is in-plane, $\theta_H=0$.  Other parameters are equal to those in Fig.~\ref{fig_disp}.
\label{fig_ddep}
}
\end{figure}

In Fig.~\ref{fig_ddep}(a), we show the extrinsic linewidth broadening $\Delta H^{(2)}$
as a function of resonant frequency for various DMI strengths.
Because of $\epsilon_{\rm DM}^2$ dependence in Eq.~(\ref{H2}), overall amplitudes of the linewidth
become larger with an increased DMI strength, whereas, for a given DMI strength, the DMI-limited scattering potential exhibits characteristic behavior
of the linewidth as a function of frequency.
Namely, with increasing the DMI strength from zero, the linewidth
changes slightly in a high frequency range while
bumps or peaks are developed in a low frequency range.
Furthermore, in a strong DMI case, for example, $D=2.3$ mJ/m$^2$
(or $\epsilon_{\rm ex}/\epsilon_{\rm DM} \sim 1.0)$ in the figure,
one can see that the linewidth is finite even at zero frequency.

We first analyze the results in the range of high frequency in Fig.~\ref{fig_ddep}(a), where
the linewidth shows $\epsilon_{\rm DM}^2$ behavior as a function of frequency.
In fact, as well as the $\epsilon_{\rm DM}^2$ dependence,
it is found that the linewidth has structured behavior resulted from
the function $\Gamma(\omega)$ as shown in Fig.~\ref{fig_ddep}(b).
For a given frequency, one can see that $\Gamma(\omega)$ increases with an increased DMI and then decreases
above a certain DMI strength.
Through detailed numerical examinations, we find that, at the turning point (call
the DMI energy $\epsilon_{\rm DM}^0$), the two lobes in equi-energy lines
illustrated in the lower panel of Fig.~\ref{fig_disp} are about to be one by
separating magnon modes at $\phi_{\bf k}=-\pi/2$ into two resonant states.
We derive $\epsilon_{\rm DM}^0$ in terms of various energies in Eq.~(\ref{app2}).
In the case of $\theta_H=0$, the DMI energy $\epsilon_{\rm DM}^0$ is approximately given by, 
\begin{equation}
\epsilon^0_{\rm DM} = \frac{ \hbar\gamma\mu_0(M_S+H_s) \epsilon_{\rm Ms} }{ 2\hbar \omega_{\rm FMR} }
\end{equation}
with $\epsilon_{\rm Ms}\!=\! \hbar\gamma \mu_0 M_S/2$.

In the case of $\epsilon_{\rm DM} \lesssim \epsilon_{\rm DM}^0$,
we find that lobes of the equi-energy line consist of small ${\bf k}$ points, namely $kd\ll 1$.
Thus, in this weak DMI regime, we find that a calculated linewidth function $\Gamma(\omega)$
is consistent with the free-DMI results by Ref. [\onlinecite{Arias}].
A detailed expression of $\Gamma(\omega)$ for the weak DMI is written in Eq.~(\ref{app_weakGamma}).
According to the equation, the increasing behavior in the weak DMI regime ($\epsilon_{\rm DM} \lesssim \epsilon_{\rm DM}^0$)
of Fig.~\ref{fig_ddep}(b) originates from an enlarged size of the equi-energy lobes in the ${\bf k}$-space
by increasing the DMI strength.  In other words, the enlarged lobes are directly related to 
more available resonant states to give rise to the increased linewidth.

In the strong DMI case of $\epsilon_{\rm DM} \gtrsim \epsilon_{\rm DM}^0$, on the other hand,
$\Gamma(\omega)$ shows a decreasing behavior as a function of DMI strength 
even though the size of equi-energy lobes becomes larger as shown in Fig.~\ref{fig_disp}(b).
In this regime, there are many resonant states with large ${\bf k}$-points, namely $kd\gtrsim 1$.
At these ${\bf k}$ points, the magnon dispersion is dominantly governed by the exchange energy, $\epsilon_{\rm ex}k^2 d^2$.
This is a rapidly increasing function of $kd$ and thus,
results in small magnon density of states or less-available resonant states.
This explanation is also consistent with the expression of $\Gamma(\omega)$ for $kd\gtrsim 1$  in Eq.~(\ref{app_strongGamma}).

\begin{figure}[t]
\centering
\includegraphics[width=45.0ex, angle=0.0]{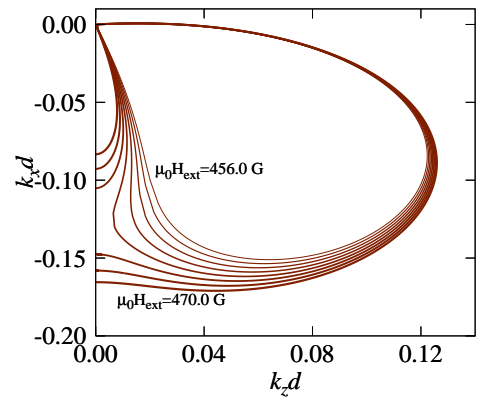}
\caption{(color online) In order to resolve the origin of the peak in Fig.~\ref{fig_ddep}(a), equi-energy lines
are plotted in the case of $D=1.5$ mJ/m$^2$ around the external magnetic field ($464.4$ G) exhibiting the peak.
We increase the magnetic field by $2.0$ G and use the same parameters as those in Fig.~\ref{fig_ddep}.
}
\label{fig_mrs}
\end{figure}

We next discuss the origin of bumps (appeared for the case $D<1.2$ mJ/m$^2$) and peaks (appeared for the case $D\geq1.2$ mJ/m$^2$) 
in Fig.~\ref{fig_ddep}(a).
We find that the bumps appear when the DMI energy equal to $\epsilon_{\rm DM}^0$.
Thus, the appearance of resonant state at $\phi_{\bf k}=-\pi/2$ is responsible for the bumps.
On the other hand, the peaks in Fig.~\ref{fig_ddep}(a) have somewhat different origins,
more complicated behavior of the energy dispersion.
At the points where the peaks are developed,
we find that there are dual solutions of the cubic equation 
derived from $\epsilon({\bf k})\!=\!\sqrt{\epsilon_{11}^0\epsilon_{22}^0}$.
For example, in the case of $D=1.5$ mJ/m$^2$, we plot equi-energy lines (the right lobe in Fig.~\ref{fig_disp})
in Fig.~\ref{fig_mrs} around the external magnetic field exhibiting the peak.
As the external magnetic field increases, the equi-energy lines are evolved to have a distorted shape, and eventually
to touch the other (the left lobe). Then, the left and right lobes are separated into outer and inner ones.
This modification of the lobes occurs at a lower magnetic field than that giving the appearance of a resonant state
at $\phi_{\bf k}=-\pi/2$.
At the magnetic field where the two lobes interact each other,
the peak is developed in the function $\Gamma(\omega)$.
This is because such formation of the lobes gives rise to a very small group velocity of
$\partial \epsilon({\bf k})/\partial k$.
In other words, this means that magnon density of states at the point is very abundant and
provides much possibility into which two magnons are scattered to increase the FMR linewidth.

As for a finite linewidth at $\omega_{\rm FMR}=0$, we find that
a DMI strength is very large, roughly $\epsilon_{\rm DM} > \sqrt{\hbar\gamma\mu_0(M_S\!+\!H_s)\epsilon_{\rm ex}}$.
Under this condition, there are lower energy states than a uniformly
magnetized state even at $H_{\rm ext}=0$, making resonant states still possible.
This can be understood by reminding the fact that spiral states are a ground state rather
than a ferromagnetic state for systems accompanying a large DMI.

\begin{figure}[ht]
\centering
\includegraphics[width=45.0ex, angle=0.0]{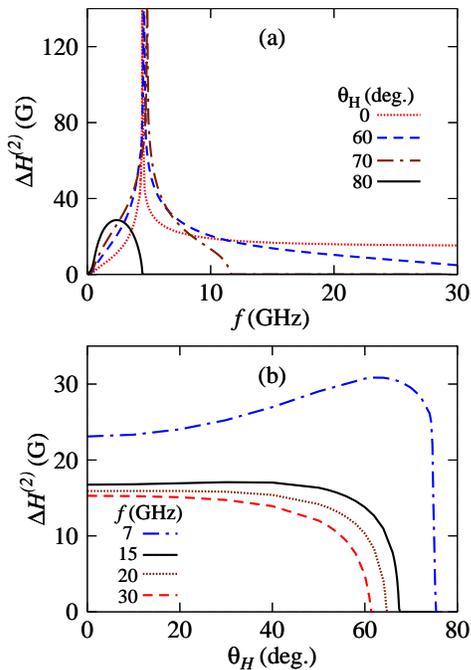}
\caption{(color online)
For a given DMI strength $D=1.5$ mJ/m$^2$, we compare
$\Delta H^{(2)}$ for different angles of applied magnetic field.
Other parameters are the same as those in Fig.~\ref{fig_disp}.
\label{fig_adep} }
\end{figure}

We now turn to the discussion about the linewidth in the out-of-plane geometry of applied magnetic fields.
Figure~\ref{fig_adep} is the linewidth for different tilted angles of the external magnetic field.
Because we consider a strong DMI case ($\epsilon_{\rm DM}\simeq \epsilon_{\rm ex}$),
the peaks appear in the linewidth curves.
We find that the peak positions are nearly unchanged with the tilting angle of magnetic field. In the range of high frequency apart from the peak structures,
there is a critical frequency above which the linewidth becomes zero, a similar behavior found in DMI-free cases~\cite{Landeros}.

In Fig.~\ref{fig_adep}(b), we plot the linewidth as a function of the tilting angle when the FMR frequency is fixed.
For a small DMI case, the critical angle above which the linewidth is zero is determined by
Eq.~(\ref{app1}) and is similar to that in the work of Ref.~[\onlinecite{Landeros}].
On the other hand, the critical angle in a strong DMI case is determined by Eq.~(\ref{app2}).

\section{Conclusion}

In this work, we have developed the theory of two-magnon scattering for a thin ferromagnetic film with DMI
as well as dipole-dipole, exchange, surface anisotropy, and Zeeman energies.
In a quantum mechanical way, we derive a expression of the FMR linewidth, taking into account scattering from
structural inhomogeneity, for example, DMI fluctuation induced by microscopic structural imperfection.
We present the extrinsic FMR linewidth in terms of various energy contributions, Eq.~(\ref{Gamma}).
In the presence of the DMI term, the magnon dispersion exhibits rich resonant
states, especially in small external magnetic fields.
Furthermore, due to the competition between the exchange and DMI energies, the resonant states are formed
at large ${\bf k}$ points (namely $k d\simeq 1$). Different from the DMI-free case ($kd\ll 1)$, it makes
difficult to derive analytical expressions from the FMR linewidth.

We find that the characteristic linewidth broadening with the DMI are twofold. 
One is the appearance of peak in the low frequency range when the DMI is strong.
This usually occurs when two lobes of the resonant states
touch each other in magnon dispersion curves.
The other is a finite value of the FMR linewidth at zero magnetic field.
This may be not possible if a uniform ferromagnetic system such as DMI-free case is the ground state.
However, in the case of strong DMI, a non-colinear spin state like spiral configuration has a lower energy than
a ferromagnetic one.
Consequently, in the presence of the strong DMI, there are still resonant states even at zero magnetic field and
thus a uniform spin state excited by microwaves in FMR experiments is
still scattered into resonant states by inhomogeneities scattering.

\section{Acknowledgements}
We acknowledge R. D. McMichael for a fruitful discussion. This work was supported by the National Research Foundation of Korea (NRF-2015M3D1A1070465, NRF-2017R1A2B2006119) and KU-KIST Graduate School of Converging Science and Technology Program.

\appendix

\section{ DMI perturbation potential }

Let's consider fluctuation from a Rashba-type spin-orbit interaction~\cite{Kim2};
\begin{eqnarray}
\Delta E_D = \frac{4\hbar\gamma}{M_S} \int d{\bf r} ~ D^d({\bf r}) \left[
 {\hat z}\cdot {\bf m}\!\times\!\frac{\partial {\bf m}}{\partial x} 
-{\hat x}\cdot {\bf m}\!\times\!\frac{\partial {\bf m}}{\partial z} \right]
\nonumber
\end{eqnarray}
where $D^d({\bf r})$ is local DMI strength for each defect.

One of feasible models for the fluctuation of this energy is
a random spatial variation of the parameter $D^d({\bf r})$
with keeping the magnon propagating vector ${\bf m}$.
This case may occur structural defects whose characteristic size is larger than
or equal to the film thickness $d$ with weak DMI.
Then, the fluctuation energy becomes, in Fourier space,
\begin{eqnarray}
	\Delta E_D &=& \sum_{{\bf k},{\bf k}'} V^d({\bf k}\!-\!{\bf k}') ik'_x d
\left[ {\bf m}(-{\bf k}) \times {\bf m}({\bf k}') \right] \cdot {\hat z} 
\nonumber\\
	&-& \sum_{{\bf k},{\bf k}'} V^d({\bf k}\!-\!{\bf k}') ik'_z d
\left[  {\bf m}(-{\bf k})\times {\bf m}({\bf k}') \right]\cdot {\hat x},
\label{fluctDMI}
\end{eqnarray}
where the defect potential energy is defined by,
\begin{eqnarray}
V^d({\bf r}) &=& \frac{4\hbar\gamma}{M_Sd} D^d({\bf r})
 =\sum_{\bf k} V^d({\bf k}) e^{i{\bf k}\cdot{\bf r}}.
\nonumber
\end{eqnarray}

As another model, we can take account of an abrupt change of the magnetization vectors,
for example, at edges of terrace on sample surfaces, near magnetic and non-magnetic impurities, etc.
Moreover, it is also known that the tilt magnetization state (spiral or skyrmion phases)
rather than a ferromagnetic state is more stable for a strong DMI strength.\cite{Iwasaki}
For those cases, we expect that magnon modes are additionally modulated
as ${\hat m}({\bf r}) e^{i({\bf k}+{\bf q})\cdot{\bf r}}$
near the defect (${\bf q}$ is a pitch vector, for example, in a spiral state).
Then, the fluctuation energy becomes similar to Eq. (\ref{fluctDMI}), but
replacing $k_x \rightarrow k_x+q_x$ and $k_z \rightarrow k_z+q_z$.
By considering small defects with a characteristic length $\lambda_c$ (order of an impurity size $\ll d$),
we further set $k_x \rightarrow 1/\lambda_c$ and $k_z \rightarrow 1/\lambda_c$ 
in Eq. (\ref{fluctDMI}).

We assume that  the defect potential energy $V^d({\bf r})$ is contributed
from atom-like and Gaussian-shaped functions located at random sites;
\begin{equation}
V^d({\bf r}) = \sum_j v_a({\bf r}-{\bf R}_j),~~~
v_a({\bf r}) = 
\frac{4\hbar \gamma D}{M_Sd} e^{-{\bf r}^2/\lambda_{\rm imp}^2}
\label{atomicP}
\end{equation}
with a defect size, $\lambda_{\rm imp}$.
Then, by inserting Eq. (\ref{atomicP}) into Eq. (\ref{fluctDMI}), the fluctuation energy becomes,
in the local coordinates,
\begin{eqnarray}
	\Delta E_D &=& 
\sum_{{\bf k},{\bf k}'}
V^d({\bf k}\!-\!{\bf k}') \frac{ i(k_x\!+\!k'_x\!+\!2/\lambda_c )d }{2} \cos\theta_M
\nonumber\\
&& {\hat x}_3\cdot \left[ {\bf m}(-{\bf k})\times {\bf m}({\bf k}')\right].
\nonumber\\
&=&\sum_{{\bf k},{\bf k}'} \sum_{\kappa,\kappa'=1,2} m_\kappa(-{\bf k})
U_{\kappa\kappa'}^{{\bf k},{\bf k}'} m_{\kappa'}({\bf k}')
\label{DMIfluct1}
\end{eqnarray}
where
\begin{eqnarray}
U_{11}^{{\bf k},{\bf k}'} &=& ~U_{22}^{{\bf k},{\bf k}'} = 0,
\nonumber\\
U_{12}^{{\bf k},{\bf k}'} &=&-U_{21}^{{\bf k}',{\bf k}} = 
\sum_j e^{-i({\bf k}\!-\!{\bf k}')\cdot {\bf R_j}} U^a({\bf k},{\bf k}').
\nonumber
\end{eqnarray}
Here, $U^a({\bf k},{\bf k}')$ is given by,
\begin{eqnarray}
U^a({\bf k},{\bf k}') = \frac{2\hbar\gamma D}{M_Sd}
f_R({\bf k}\!-\!{\bf k}')
{\cal W}({\bf k},{\bf k}') \cos\theta_M
\end{eqnarray}
where
${\cal W}({\bf k},{\bf k}') = i (k'_x+k_x) d/2$ for $\lambda_c \gtrsim d$, whereas ${\cal W} = d/\lambda_c$ for $\lambda_c \ll d$. Here, $f_R({\bf k})= e^{-k^2\lambda_{\rm imp}^2} \pi\lambda_{\rm imp}^2/L^2$
is a form factor of the atomic potential
($L$ is a side length of the ferromagnetic film).

\vspace{1cm}
\section{Critical angle of external magnetic fields}

Above a certain angle of external magnetic field, the linewidth of FMR becomes zero.
The critical angle is determined by demanding zero resonant states, or tiny size of lobes, in 
the FMR condition of $\epsilon({\bf k}) = \hbar\omega_{\rm FMR}$.
By requiring a small value of ${\bf k}$, it is straightforward to find
\begin{eqnarray}
k_R(\phi_{\bf k})d = \frac{ \left(\frac{C_1}{2C_2}\right)^2+\frac{C_0}{C_2}-\left(\frac{C_1}{2C_2}+\sin\phi_{\bf k}\right)^2}{ D_2/C_2 }
\label{smallkR}
\end{eqnarray}
where 
\begin{eqnarray}
C_0 &=& \epsilon_{\rm Ms}\left[ \epsilon_{22}^{0}\cos 2\theta_M-(\epsilon_{11}^{0}+\epsilon_{22}^{0})\sin^2\theta_M\right]
\nonumber\\
C_1 &=& 2 \sqrt{\epsilon_{11}^0\epsilon_{22}^0} \epsilon_{\rm DM}\cos\theta_M,
\nonumber\\
C_2 &=& \epsilon_{\rm Ms} \left[ \epsilon_{11}^{0}-\epsilon_{22}^{0}\sin^2\theta_M\right],
\nonumber\\
D_2 &=& (\epsilon_{11}^0\!+\!\epsilon_{22}^0)\epsilon_{\rm ex}
+\epsilon_{\rm Ms}^2 \sin^2\theta_M (\sin^2\theta_M\!-\!\cos2\theta_M )
\nonumber\\
&-&\frac{1}{2} \sin^2\phi_{\bf k} \cos^2\theta_M[2\epsilon_{\rm DM}^2+\epsilon_{\rm Ms}^2(1\!+\!\cos2\theta_M)].
\nonumber
\end{eqnarray}
Thus, for a small DMI of $\frac{C_1}{2C_2} \leq 1$, the critical angle is determined by solving the equation,
\begin{equation}
\left(\frac{C_1}{2C_2}\right)^2+\frac{C_0}{C_2}=0
\label{app1}
\end{equation}
while, $\frac{C_1}{2C_2}>1$,
\begin{equation}
\left(\frac{C_1}{2C_2}\right)^2+\frac{C_0}{C_2}-\left(\frac{C_1}{2C_2}-1\right)^2=0.
\label{app2}
\end{equation}

\section{ Evaluation of the linewidth function $\Gamma(\omega)$ }

The linewidth function $\Gamma(\omega_{\rm FMR})$  can be rewritten as, from Eq.~(\ref{Gamma}),
\begin{flalign}
\Gamma&(\omega)=
\frac{1}{\cos(\theta_M\!-\!\theta_H)}
\frac{\epsilon_{\rm ex} }{ (\epsilon_{22}^0\!+\!\epsilon^0_{11}) }
\frac{\pi d^2}{L^2} \sum_{{\bf k}} f_c(k)
\nonumber\\
&
\frac{\epsilon_{11}^0\epsilon_{11}({\bf k})+\epsilon_{22}^0\epsilon_{22}({\bf k}) }{
\sqrt{ 4\epsilon_{11}({\bf k})\epsilon_{22}({\bf k})\!-\![\epsilon_{12}({\bf k})+\epsilon_{21}({\bf k})]^2} }
\delta(\hbar\omega\!-\!\epsilon({\bf k})).
\end{flalign}

For a weak DMI of $\frac{C_1}{2C_2} \leq 1$ and due to $kd\ll 1$, 
the delta function in the equation can be approximated by,
\[
\delta(\hbar\omega\!-\!\epsilon({\bf k})) \simeq \frac{2\sqrt{\epsilon_{11}^0\epsilon_{22}^0}}{|C_0|d}
\delta(k-k_{R}(\phi_{\bf k}))
\]
from Eq.~(\ref{smallkR}) and then, $\Gamma(\omega)$ is further simplified to
\begin{flalign}
\Gamma&(\omega) \simeq
\frac{ \epsilon_{\rm ex} ([\epsilon_{22}^{0}]^2\!+\![\epsilon^{0}_{11}]^2) }
{2\pi \cos(\theta_M\!-\!\theta_H) (\epsilon_{22}^0\!+\!\epsilon^0_{11})|C_0| }
\int_{\phi_a}^{\phi_b} d\phi k_R(\phi) d
\label{app_weakGamma}
\end{flalign}
where
\begin{eqnarray}
\sin\phi_a &=&
-\frac{C_1}{2C_2} -\sqrt{\left(\frac{C_1}{2C_2}\right)^2+\frac{C_0}{C_2}} 
\nonumber\\
\sin\phi_b &=&
-\frac{C_1}{2C_2}+ \sqrt{\left(\frac{C_1}{2C_2}\right)^2+\frac{C_0}{C_2}}.
\nonumber
\end{eqnarray}

Now we consider a strong DMI of $\frac{C_1}{2C_2} \gg 1$ where the dipole energy $\epsilon_{\rm Ms}$ is relatively unimportant.
In this case, eigenenergy of magnon can be approximated by, from Eq.~(\ref{eigenenergy}),
\[
\epsilon({\bf k}) = {\rm Im}[\epsilon_{12}({\bf k})]+\sqrt{
(\epsilon_{11}^0+\epsilon_{\rm ex} k^2 d^2)
(\epsilon_{22}^0+\epsilon_{\rm ex} k^2 d^2)}.
\]
By solving this, one can show that the resonant states appear at,
\begin{equation}
k_R(\phi_{\bf k})d \simeq \frac{\epsilon_{\rm DM} |\sin\phi_{\bf k}|\cos\theta_M}{\epsilon_{\rm ex}},
~~~-\frac{\pi}{2}\leq \phi_{\bf k} \leq 0.
\label{largekR}
\end{equation}
Then, magnon density of states at a ${\bf k}$ point is given  by,
\[
\delta(\hbar\omega\!-\!\epsilon({\bf k})) \simeq 
\frac{1}{ \epsilon_{11}^0\!+\!\epsilon_{22}^0\!-\!\hbar\omega+2 \frac{{\tilde\epsilon}_{\rm DM}^2}{\epsilon_{\rm ex}}}
\delta(kd-k_{R}(\phi_{\bf k})d)
\]
where we abbreviate ${\tilde\epsilon}_{\rm DM}=\epsilon_{\rm DM} |\sin\phi_{\bf k}|\cos\theta_M$,
and the associated linewidth function $\Gamma(\omega)$ becomes
\begin{flalign}
\Gamma(\omega) &\simeq
\frac{ 1}{4\pi \cos(\theta_M\!-\!\theta_H) }
\nonumber\\
&\int_{-\pi/2}^0 d\phi_{\bf k}
\frac{\epsilon_{\rm ex}}{\epsilon_{11}^0\!+\!\epsilon_{22}^0\!-\!\hbar\omega
+2\frac{{\tilde\epsilon}_{\rm DM}^2}{\epsilon_{\rm ex}} }.
\label{app_strongGamma}
\end{flalign}


\end{document}